\documentclass[aps,twocolumn]{revtex4}
\usepackage{graphicx}

\begin{document}
\title{A 1 W injection locked cw titanium:sapphire laser}

\author{E. A. Cummings, M. S. Hicken, and S. D. Bergeson}
\affiliation{Department of Physics and Astronomy, Brigham Young
University, Provo, UT  84602}

\begin{abstract}
We report an injection-locked cw titanium:sapphire ring laser at
846 nm.  It produces 1.00 W in a single frequency when pumped with
5.5 W.  Single frequency operation requires only a few milliwatts
of injected power.
\end{abstract}


\maketitle

Single frequency cw lasers are central to many experiments in
spectroscopy, atomic physics, nonlinear optics, quantum optics,
metrology, ranging, communications, and other fields.  A number of
tunable and fixed-frequency lasers are commercially available to
cover the near ultraviolet, visible, and near infrared spectral
regions.  Dye lasers, Ti:sapphire lasers, optical parametric
oscillators, Nd:YAG lasers, various gas lasers, diode lasers, and
many others regularly find applications in these fields.

A few years ago, high-power single frequency diode lasers began
to compete over selected wavelength ranges with moderate power
Ti:sapphire lasers and dye lasers (see, for example,
\cite{busse93,praeger98,wilson98,ferrari99,oates99} and many
others). These diode lasers were relatively inexpensive and could
be frequency-stabilized, either by injection locking or by an
external cavity, making them ideal for certain classes of
experiments.  However, many of these high-power single frequency
diode lasers have become increasingly expensive and difficult to
obtain at particular wavelengths.

In this paper we describe a mod\-er\-ate pow\-er, cw, tun\-a\-ble,
sin\-gle-fre\-quen\-cy in\-jec\-tion-locked Ti:sapphire laser.
Injection-locking a high power laser with a low-power master
laser reproduces the master laser at higher power with good
fidelity \cite{siegman86,farinas95}.  Injection-locking has been
demonstrated in a wide range of cw laser systems, including
Nd:YAG \cite{peng85,nabors89,teehan00}, argon-ion \cite{man84},
He-Ne \cite{urisu81}, diode \cite{goldberg87}, and dye lasers
\cite{couillaud84}. It has also been demonstrated for pulsed
Ti:sapphire \cite{bair88,ni98} and dye lasers
\cite{eikema97,bergeson98}. However, it has apparently not been
demonstrated in cw Ti:sapphire lasers.

The injection laser determines the output wavelength and forces
both single frequency and uni-directional operation.  It
eliminates birefringent filters and optical diodes in more
traditional Ti:sapphire lasers
\cite{albers86,schulz88,harrison91,zimmermann95}.  The wavelength
tunability of our system mirrors the tunability of the diode
laser. While no single diode laser covers the entire tuning range
of Ti:sapphire, diodes are available at many wavelengths from
0.66 to 1.1 $\mu$m.  Furthermore, many kinds of experiments use
only a limited wavelength range---some using only several GHz of
tuning---and our laser is well-suited for those experiments.

\begin{figure}
\centerline{\rotatebox{270}{\scalebox{.35}{\includegraphics{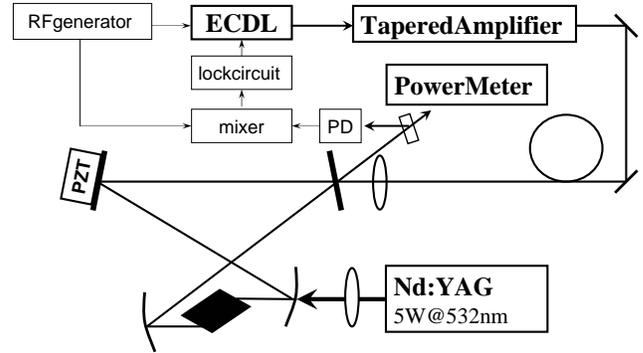}}}}
\caption{A schematic diagram of the experimental layout.
ECDL=extended cavity diode laser, PD=photodiode,
PZT=piezoelectric transducer. \label{fig:layout}}
\end{figure}

Figure \ref{fig:layout} shows a schematic diagram of our
experiment.  The master laser is an extended-cavity diode laser at
846 nm \cite{newfocus}, amplified in a single pass through a
tapered laser diode \cite{sdl}.  It is optically isolated and
coupled into a single-mode optical fiber. It delivers up to 30 mW
to the Ti:sapphire laser cavity.

A pair of lenses mode-matches the collimated output from the fiber
into the Ti:sapphire ring cavity. The coupling efficiency is
typically 75\% into the TEM$_{00}$ Gaussian mode of the cavity,
and higher order modes are about 1\% or less than the Gaussian
mode. The Ti:sapphire cavity is a four-mirror folded (bowtie)
cavity around a Ti:sapphire crystal.  The crystal is 10 mm long,
3 mm diameter, Brewster cut, with a low-power single-pass
absorption coefficient of 2.1 at 532 nm.  It is mounted in a
water-cooled brass housing.  The cavity's flat output coupler
reflectivity is 96.6\%. The reflectivities of the other three
mirrors are all $> 99.5$\%.  The radius of curvature for the
curved mirrors is 100 mm.  The short distance between the two
curved mirrors, including the path through the crystal, is 114
mm.  The long distance between the two curved mirrors is 1020
mm.  The angle of incidence on the curved mirrors is
8$^{\mbox{\footnotesize{o}}}$.  These distances and angle of
incidence are chosen to compensate for the astigmatism introduced
by the Brewster-cut crystal \cite{kogelnik72}. The
``cold-cavity'' finesse, measured without pumping the crystal, is
110.

Up to 5.5 W from the 532 nm pump laser is focused into the middle
of the Ti:sapphire crystal.  Not shown in the Figure is a
telescope in the green laser beam, with one lens mounted on a
micrometer stage, to optimize the focus of the green laser beam
into the crystal.

The diode current in the master laser is modulated at 37.15 MHz.
This produces the frequency sidebands necessary to lock the
master laser to the Ti:sapphire cavity using the Pound-Drever-Hall
technique \cite{pound83}. The electronic feedback circuit is a
two-stage integrator, with fast feedback to the master laser
current, and slow feedback to the master laser cavity length. One
of the Ti:sapphire cavity mirrors is mounted on a piezo-electric
crystal, allowing approximately 8 GHz of continuous frequency
tuning.  To achieve a particular wavelength from the Ti:sapphire
laser, it is necessary to first set the approximate wavelength of
the diode laser before injection-locking the Ti:sapphire laser.
Fine wavelength tuning is achieved by scanning the Ti:sapphire
cavity after engaging the lock.  The optical signal used for the
feedback circuit comes from a weak reflection from an uncoated
quartz optical flat in the high power output from the Ti:sapphire
laser.  This reflected beam is attenuated to prevent saturation
in the feedback photodiode.  All of the power measurements
reported here are made after the quartz flat, as shown in Figure
\ref{fig:layout}.

\begin{figure}
\centerline{\rotatebox{270}{\scalebox{.35}{\includegraphics{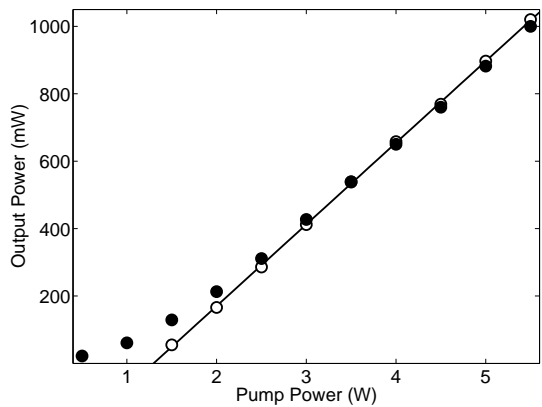}}}}
\caption{Input-output data for the Ti:sapphire laser for
injection-locked ($\bullet$) and free-running ($\circ$)
operation.  The straight lines show the fitted threshold power
and efficiency.  The slope efficiency is 24\%. \label{fig:gain1}}
\end{figure}

Figure \ref{fig:gain1} shows a plot of the power out of the
injection-locked laser as a function of pump power.  It also
shows the free-running laser output power, when the injection
laser is blocked.  The injection-locked and free-running lasers
have essentially the same maximum power output and slope
efficiency.  However, the injection-locked laser has a
significantly lower threshold power. Without injection-locking,
the laser does not oscillate until photons are spontaneously
emitted into the cavity mode.  However, the injection laser puts
photons into the cavity mode, and the laser oscillates at pump
powers below the non-injected threshold.

Our measured slope efficiency (23\%) matches calculations based on
a theoretical model \cite{harrison91}.  However the threshold
power (about 1.2 W) is considerably higher than the calculation
(about 0.3 W).  This implies that either the cavity waist or the
pump laser waist (or both) are not as small as they should be.
This is probably due to thermal lensing and to a mismatch between
the confocal parameter for the pump laser beam and the crystal
length.

\begin{figure}
\centerline{\rotatebox{270}{\scalebox{.35}{\includegraphics{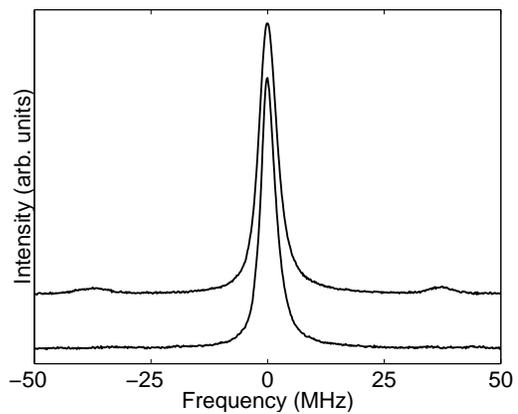}}}}
\caption{Optical frequency spectrum of the laser.  The top trace
shows the spectrum of the injection laser.  The bottom trace
shows the spectrum of the amplified laser.  The traces are offset
vertically for clarity.  \label{fig:spectrum}}
\end{figure}

When properly injected, the Ti:sapphire laser output is a single
frequency.  We monitor the frequency spectrum of the Ti:sapphire
laser output with a scanning Fabry-Perot cavity (free spectral
range = 2 GHz, finesse = 400).  The optical spectrum of the
injection laser beam is shown in Figure \ref{fig:spectrum} (top
trace). Also shown is the optical spectrum of the amplified beam
(bottom trace).  In both cases, the measured spectral width of
the laser reflects the limit of resolution of the scanning
Fabry-Perot cavity.

\begin{figure}
\centerline{\rotatebox{270}{\scalebox{.35}{\includegraphics{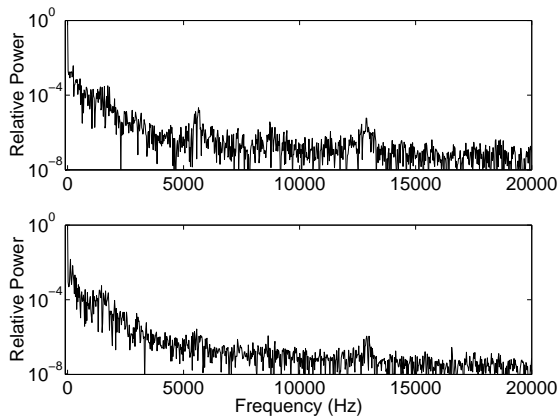}}}}
\caption{Power spectrum of the relative optical frequency between
the Fabry-Perot cavity and the laser.  The top trace compares the
cavity to the injection laser.  The bottom trace compares the
cavity to the amplified Ti:sapphire laser.
\label{fig:powerspectrum}}
\end{figure}

While the scanning Fabry-Perot does not have enough resolution to
measure the spectral width of the laser, it can still give us an
indication of the frequency content of the beam.  Taking the
Fabry-Perot cavity out of its sweeping-mode operation, we tune it
into resonance with the Ti:sapphire laser.  The signal output
from the Fabry-Perot cavity is now a measure of the relative
frequency of the cavity and the laser.  The power spectrum of
this measurement is shown in Figure \ref{fig:powerspectrum}.  The
upper trace in the Figure compares the cavity to the injection
laser. The lower trace compares the cavity to the Ti:sapphire
laser at an output power near 1 W.  The two traces are
essentially identical.  They both fall rapidly in the first
hundred Hz, more slowly to 5 kHz, and are flat at higher
frequencies.

The minimum power required to maintain single frequency operation
apparently depends critically on the feedback circuit.  For our
laser with a 1134 mm round trip cavity length, the minimum power,
P$_{\mbox{\footnotesize{min}}}$, for different pump powers is
shown in Table \ref{tab:minpower}.  As the injection power
decreases, the amplitude noise in the output increases
(indicating an improper feedback gain) long before the frequency
noise increases.  We also made smaller amplifier cavity, with 50
mm radius of curvature mirrors and an overall cavity length of
360 mm. The smaller cavity, which has the same waist size,
operates similarly to the system described in this paper in both
threshold power and slope efficiency.  However, the minimum power
required to maintain single frequency operation is lower by a
factor of 3.

\begin{table}
\caption{Minimum injection power, P$_{\mbox{\footnotesize{min}}}$,
and output power, P$_{\mbox{\footnotesize{out}}}$, for different
pump powers. \label{tab:minpower}}
\begin{center}
\begin{tabular}{ccc}
\hline \hline Pump Power (W) & P$_{\mbox{\footnotesize{min}}}$
(mW) & P$_{\mbox{\footnotesize{out}}}$ (W) \\
\hline 2.0 & 3 & 0.22 \\
3.0 & 6 & 0.49 \\
4.0 & 12 & 0.74 \\
5.0 & 15 & 1.00 \\
\hline
\end{tabular}
\end{center}
\end{table}

We have demonstrated a single frequency cw injection-locked
Ti:sapphire laser at 846 nm, continuously tunable over an 8 GHz
range, with an output power of 1 W.  The wavelength tuning range
is determined by the tuning range of the laser diode used to
inject the Ti:sapphire laser.  A broad range of laser diodes are
available that cover much of the Ti:sapphire gain region.  At
those wavelengths, an injection-locked Ti:sapphire laser will
produce output powers in the 1 W range at relatively low cost.

This work is supported in part by grants from the Research
Corporation and from the National Science Foundation under Grant
No. PHY-9985027.


\begin{thebibliography}{}
\bibitem{sdl}
The tapered amplifier is a Model 8613 laser from SDL (80 Rose
Orchard Way, San Jose, CA 95134-1365) with the rear cavity optics
removed, configured as a single-pass amplifier.
\end{thebibliography}
\end{document}